\documentclass[pra,twocolumn,amsfonts]{revtex4}
\usepackage{graphicx}
\usepackage{amsmath}
\usepackage{amsfonts}
\usepackage[dvips]{epsfig}
\usepackage{bm}
\usepackage{epsf}

\def\pra{ Phys. Rev. A}
\def\prl{ Phys. Rev. Lett.}

\begin{document}
\author{S. Marcovitch$^1$, A. Retzker$^2$, M. B. Plenio$^2$ and
B. Reznik$^1$}
\title{Critical and noncritical long range entanglement in Klein-Gordon fields}
\affiliation{$^1$ School of Physics and Astronomy,
Raymond and Beverly Sackler Faculty of Exact Sciences,
Tel-Aviv University, Tel-Aviv 69978, Israel\\
$^2$ Institute for Mathematical Sciences, Imperial College London,
London SW7 2PE, United Kingdom, \\and QOLS, The Blackett Laboratory,
Imperial College London, Prince Consort Road, London SW7 2BW,
United Kingdom}
\begin{abstract}
We investigate the entanglement between two spatially separated
intervals in the vacuum state of a free $1D$ Klein-Gordon field
by means of explicit computations in the continuum limit of the
linear harmonic chain. We demonstrate that the entanglement, which
we quantify by the logarithmic negativity, is finite with no further
need for renormalization. We find that the quantum correlations
are scale-invariant and are determined by a function depending
on the ratio of distance to length only. They decay much faster
than the classical correlations as in the critical limit long
range entanglement decays exponentially for separations larger
than the size of the blocks, while classical
correlations follow a power law decay.
With decreasing distance of the blocks, the entanglement diverges
as a power law in the distance. The noncritical regime manifests
richer behavior, as the entanglement depends both on the size of
the blocks and on their separation. In correspondence with the
von Neumann entropy also long-range entanglement distinguishes
critical from noncritical systems.
\end{abstract}
\maketitle
\date{\today}

The scaling of block entanglement in both harmonic and spin chains
has received considerable attention recently \cite{EisertCP08}.
The scaling of the entanglement entropy $S(\rho_A)=-tr[\rho_A\log\rho_A]$
of a block $A$ has been found to behave in a universal way in
one-dimensional
critical systems. Explicit computations for the Klein-Gordon
massless field using density matrices have shown that the
entanglement entropy is proportional to the area of the boundary
\cite{bombelli,srednicki}. Using general conformal field theory
methods it has been shown that the entropy of a block with size
$l$ scales as $(\bar{c}/3)\log l$ for bosonic fields
\cite{callan,kabat,holzhey,dowker}, where $\bar{c}=1$ is the
central charge in the one-dimensional case. This result has been
verified analytically \cite{korepin,korepin1} using quantum
information methods both in critical bosonic spin chains and
in critical linear harmonic chains (HC) \cite{modewise2}, where
the area-law has been proven in higher dimensions in
\cite{plenio2007,plenio2002}. In noncritical chains however the
entropy saturates for blocks larger than the correlation length
$\xi\sim m^{-1}$,
where $m$ is the energy gap, given in dimensionless units where
$\hbar=c=1$. 
The von Neumann entropy however,
requires further renormalization as it diverges in the continuum
limit. Due to this divergence the mutual information $I=S(\rho_A)
+S(\rho_B)-S(\rho_{AB})$ of two regions $A$ and $B$ has been
suggested as a better measure \cite{finite_mutual} since it admits
a finite value in the continuum limit. For critical
fields $I$ can still be computed using conformal field theory techniques,
where it scales as a power law with the separation between the regions,
and decays exponentially in the massive case \cite{casini}.
However, mutual information is not a genuine measure of entanglement
as it includes both classical and quantum correlations as
demonstrated by the fact that it does not vanish for separable
states \cite{PlenioV07}. The mutual information is an upper bound
to many other entanglement measures, as, for example, the
distillable entanglement and the relative entropy of entanglement
\cite{PlenioV07}. Hence we will consider in the following the
logarithmic negativity $E_{LN}$ as a quantifier of entanglement
\cite{negativity} for both pure and mixed states, due in part to
the relative ease with which it is computed.

We present numerical evidence that the logarithmic negativity
admits a finite value in the critical and non critical field limit,
making it a promising candidate for studying entanglement of
quantum fields. We study the scaling of entanglement between
two spatially separated blocks in a free one-dimensional
Klein-Gordon field. Moreover a non-zero value of the logarithmic
negativity implies distillability \cite{giedke} in Gaussian
systems. At the moment, to the best of our knowledge, there
are no methods from conformal field theory, which enable one
to compute the logarithmic negativity analytically, and we
thus obtain numerically its asymptotic behaviors.

In the following we summarize our main findings before presenting
the numerical analysis. Firstly, as there are no length scales in the
critical field, any well-defined finite physical property must
depend only on a single parameter $r\equiv d/l$, where $l$ is
the length of each of the blocks and $d$ is their separation.
This property is automatically valid both for the mutual
information and the logarithmic negativity both being finite in the continuum limit, due to the
following reason. When studying the critical case the continuum
limit is taken by taking the lattice constant to zero and the
coupling coefficient to infinity in a way that the propagation
velocity remains constant. Both the negativity and the mutual
information do not depend on the coupling and thus the continuum
limit is taken just by increasing the number of oscillators. Thus
only the ratio $d/l$ can enter as a parameter in the continuum limit.

Secondly, while the classical correlations between two sites
or blocks decay as a power law in the critical regime, we
find that quantum correlations measured by $E_{LN}$ decay exponentially
with the distance in this regime following
$E_{LN}\sim e^{-\beta_c r}$ for $r> 1/2$, where $\beta_c\sim 2\sqrt{2}$
with accuracy better than $1\%$. This result improves a previously
found lower bound, $E_{LN}\sim e^{-r^2}$ for bosons \cite{silman} and fermions \cite{silman2}.
As the blocks approach each other, {\it i.e.} $r\to 0$, we find that the
$E_{LN}$ diverges as a power law $r^{-\alpha}$, where $\alpha=
\frac{1}{3}$ with accuracy better than $1\%$. Both the mutual
information $I$ and the logarithmic negativity $E_{LN}$ are upper
bounds to the distillable entanglement. Here we observe that
$E_{LN}$ is much tighter though, since $I$ scales
as a power law throughout, where we obtained numerically $I\sim r^{-0.05}$.
Hence classical correlations exhibit a power law scaling
while quantum correlations exhibit an exponential decay.


Thirdly, since the noncritical chain has a length scale
proportional to the inverse mass, this system is characterized
by two dimensionless parameters $d\to d \xi^{-1}$ and $l\to l \xi^{-1}$,
such that $m=1$, where the critical behavior is obtained in the
limit where $d\to0$ and $l\to0$. We find that just like the block
entropy, long range entanglement allows us to discriminate between
critical and noncritical fields. With respect to the blocks' size
$l$, $E_{LN}(l)$ saturates with increasing $l$, while it diverges
for the critical field. Interestingly, the saturation occurs for
$l>l_s\sim d+1$, in contrast to the block entropy for
which the saturation occurs for $l>l_s=1$ ($m=1$). With respect to
the blocks' separation $E_{LN}(d)\sim \exp(-\beta_{nc}(l)\, d^2)$ for
$d>l>1$. As $d\to 0$ the entanglement in both critical and
noncritical fields exhibits a similar behaviour since in the
noncritical regime (finite $l$), $E_{LN}$ diverges as a
power law as well.

Entanglement between groups of discrete sites has been discussed
before in various setups, such as the (discrete) Bose-Hubbard model
\cite{hubbard}, spin chains \cite{keating} and the ion trap \cite{retzker}. Our work, on the
other hand, studies the behavior of long-range entanglement in
continuous fields.

Let us start by describing the correspondence between a continuous
Klein-Gordon field and the discrete chain, and review the computation
of several quantum information measures. The free one-dimensional
Klein-Gordon Hamiltonian, $H=\int \mathcal{H}dx$, where
\begin{equation}
\mathcal{H}=\frac{1}{2}\pi^2+\frac{1}{2}(\nabla\phi)^2+\frac{1}{2}m^2\phi^2
\label{klein}
\end{equation}
corresponds upon discretization with a spacing $a$ to
\begin{equation}
        H=\frac{1}{2}\sum_{i=-\infty}^{\infty}
        \left(a\,\pi_i^2+\frac{1}{a}\left(\phi_i-\phi_{i-1}\right)^2+
        a \,m^2 \phi_i^2\right).
\label{discrete}
\end{equation}
Substituting $\pi_i \to p_i$ and $\phi_i \to q_i$, 
transforming to circular boundary conditions and writing
in dimensionless form we find 
\begin{equation}
        H = \frac{1}{2}\sum_{n=1}^N
        \left(q_n^2+p_n^2-\alpha q_n q_{n+1}\right),
\label{hc}
\end{equation}
where $q_n$ and $p_n$ are canonical variables ($q_1=q_{N+1}$), $N$ is the
number of oscillators in the chain and $0<\alpha<1$ is the
coupling constant. The correlation length $\xi$ in units of
the oscillators spacing is defined as
\begin{equation}
        \xi=\sqrt{\frac{1}{2(1-\alpha)}}.
\label{Nt}
\end{equation}
The continuum limit of the harmonic chain corresponds to Eq.
(\ref{klein}) in the strong coupling limit, $\alpha\to 1$,
given that $N\to\infty$ and $m=N/\xi$ is kept constant to
ensure a constant propagation speed. The system is critical,
{\it i.e.} $m\to0$, when $N\ll\xi$.

The spectrum
of Eq. (\ref{hc}) is given by
\begin{equation}
        \nu_k=\sqrt{1-\alpha\cos\theta_k},
\end{equation}
where $\theta_k=2\pi k/N$ and $k=0,1,\dots,N $.
Then we can express
\begin{equation}
\label{fields}
\begin{split}
&q_n=\frac{1}{\sqrt{N}}\sum_k\frac{1}{\sqrt{2\nu_k}}[a_k e^{i \theta_k n t}+\rm{H.C}],
\\& p_n=\frac{-i}{\sqrt{N}}\sum_k\frac{\nu_k}{\sqrt{2}}[a_k e^{-i \theta_k n t}-\rm{H.C}],
\end{split}
\end{equation}
where $[a_k,a_k^\dagger]=1$, and
the two-point vacuum correlation matrices $G$ and $H$ are:
\begin{equation}
\begin{split}
&G_{ij}=\langle0|q_iq_j|0\rangle=g_{(i-j)},
H_{ij}=\langle0|p_ip_j|0\rangle=h_{(i-j)}.
\end{split}
\end{equation}

Throughout this paper we consider two separated blocks $A$ and
$B$ with the same size $l=\xi L$ and separation $d=\xi D$, where
$L$ and $D$ are the number of oscillators in the blocks and their
separation, respectively.

For clarity we show how the von Neumann entropy $S$, the mutual
information $I$ and the logarithmic negativity $E_{LN}$ \cite{negativity}
may be computed efficiently for Gaussian states
\begin{equation}
\label{von}
        S(A)=\sum_{j}
        \left( f(\lambda_j+1/2)-f(\lambda_j-1/2) \right),
\end{equation}
where $f(x)=x\log x$ and $\lambda_j$ are the eigenvalues of
$iG_A H_A$ with $G_A$ being the restriction of $G$ to a block
$A$. For the logarithmic negativity we find
\begin{equation}
\label{logneg}
        E_{LN}=-\sum_j \log_2\big(\rm{min}\{2\tilde{\lambda}_j,1\}\big),
\end{equation}
where $\tilde{\lambda}_j$ are the eigenvalues of $iG_{A\cup B}
\tilde H_{A\cup B}$, where $\tilde H_{A\cup B}$ is obtained from
$H_{A\cup B}$ by time-reversal in $B$, {\it i.e.} $p_B\to-p_B$.

In the continuous limit of the critical field the two-point
correlation functions $g(x_1-x_2),\,h(x_1-x_2)$ are
\begin{equation}
\begin{split}
        g(x)\sim g_0-\log|x| \; \mbox{ and }\;
        h(x)\sim -\frac{1}{x^2}.
\end{split}
\end{equation}

The two-point correlation functions in the noncritical case depend on the mass $m$,
where in the asymptotic limit, $x>>m^{-1}$
\begin{equation}
\begin{split}
        g(x)\sim -\frac{e^{-|x|/m}}{\sqrt{|x|}}\;
        \mbox{ and }\;
        h(x)\sim \frac{e^{-|x|/m}}{\sqrt{|x|^3}}.
\end{split}
\end{equation}

\begin{figure}[ht]
\center{
\includegraphics[width=3.4in]{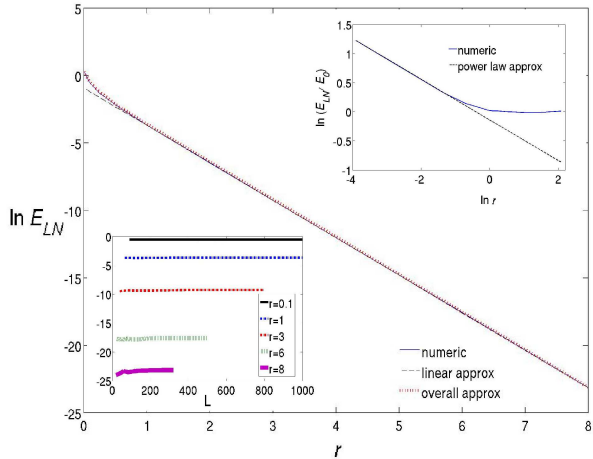}
\caption
{(Color online). Critical HC: $\ln E_{LN}$ as a function of
$r\equiv d/l$. For $r>0.5$ the linear approximation practically
coincides with the computed values, $\beta_c\sim 2\sqrt{2}$.
The dotted line is the the overall estimation, (Eq. \ref{overall}).
Upper inset: $\ln(E_1/E_0)$ as a function of $\ln r$. Lower inset:
$\ln E_{LN}$ as a function of $L$ for different values of $r$
($D$, the number of oscillators that separate the blocks, increases
with $L$).}
\label{hc_bc}
}
\end{figure}

We proceed with the presentation of the numerical results.
We examine large chains with $N=2\cdot10^4$ (and $N=4\cdot10^4$
oscillators in order to confirm the continuum limit). We begin
with the critical regime where we take $\alpha=1-10^{-12}$
(deep in the critical limit). In figure \ref{hc_bc} we present
$\ln E_{LN}$ as a function of $r$. For $r> 0.5$ the linear
approximation practically coincides with the computed values, 
$E_{LN}(r>0.5)=E_0\sim e^{-\beta_c r}$,
where the obtained constant is $\beta_c\sim 2\sqrt{2}$ to $1\%$
accuracy. 


In the upper inset we observe on a log-log scale the power
law correction to the exponential approximation. Assuming $E_{LN}=E_1\sim r^{-\alpha}
\,e^{-\beta_c r}$, we find $\ln(E_1/E_0)\sim-\alpha\, \ln r$.
For $r<0.25$ ($\ln r<-1.4$) we obtain numerically $\alpha=\frac{1}{3}$
to $1\%$ accuracy (this number was also numerically observed
for critical spin systems in \cite{sugato}). Note that $\alpha$
is identical to the prefactor of the entanglement entropy $S(l)$
in the critical field.

In the lower inset we confirm that $E_{LN}$ is scale invariant,
depending only on $d/l$ in the critical limit. We plot $\ln E_{LN}$
as a function of $L$, the number of oscillators in each of the
blocks, such that the ratio $r\equiv D/L$ is kept constant. The
plots are given for different values of $r$. The curves are
approximately constant for $L$ sufficiently large to correspond
to the continuum limit. We have also verified the same scale
invariance for the mutual information.

For arbitrary values of $r$ we find
\begin{equation}
        E_{LN}^{\rm{critical}} \sim
        \left(a r^{-\alpha}+f(r)\right)e^{-\beta_c r},
\label{overall}
\end{equation}
where $f(r)\sim e^{-\gamma/r}$. Note that as expected $f(r\gg1)\to1$
and $f(r\to0)\to0$. (Numerically we obtain $\gamma\sim 3/2$ and
$a\sim 4/3$.) The dotted line in figure \ref{hc_bc} shows Eq.
(\ref{overall}) (on logarithmic scale),
and provides a very good approximation.

\begin{figure}[ht]
\center{
\includegraphics[width=3.4in]{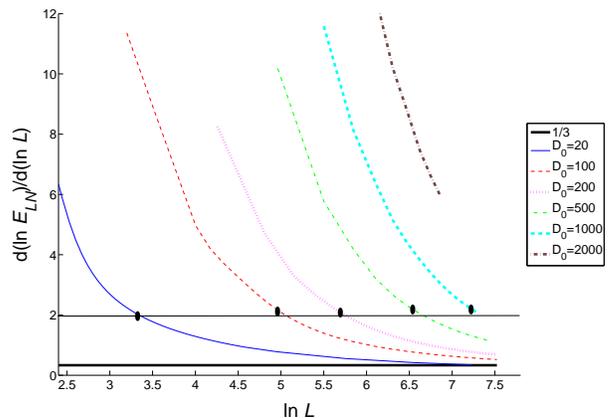}
\caption{(Color online). Critical HC: $d(\ln E_{LN})/d(\ln L)$ as a function of $L$
for several values of the separation $D_0$: $20,100,200,1000,2000$.
The dots indicate the saddle points where the power is $\sim2$.
$E_{LN}(L/D_0\to0)\sim e^L$, where $E_{LN}(L/D_0\gg1)\sim L^{1/3}$
(seen for $D_0=20$).
}
\label{hc_evidence2}
}
\end{figure}

Let us now analyze Eq. (\ref{overall}) with respect to the blocks'
size $L$, keeping their separation $D_0$ constant. First we note
that the first order exponential term $E_0(l)\sim {\rm{exp}}
(-\beta_c d_0/l)$ has a saddle point $d^2 E_0/d l^2=0$ at
$l=\beta_c d_0/2$,
in which the scaling turns from exponential at $l \to 0$ to a
power of $2$. At $l\sim \beta_c d_0$, $E_0(l)$ already scales
logarithmically and for $l/d_0\gg1$, $E_0(l)$ saturates. However,
at this limit the power law correction becomes the dominant factor,
where $E_{LN}\sim l^{1/3}$.
As the power law is obtained from the slope in a log-log plot, we show
in figure \ref{hc_evidence2}, $d(\ln E_{LN})/d(\ln L)$ as a function of $\ln L$
for several values of the separations $D_0$.
We also add the saddle points at $L\sim \sqrt{2} D_0$ for each of the curves, which
indicate the power of $2$.
In addition, asymptotically the plots tend to the $1/3$ power.

\begin{figure}[ht]
\center{
\includegraphics[width=3.4in]{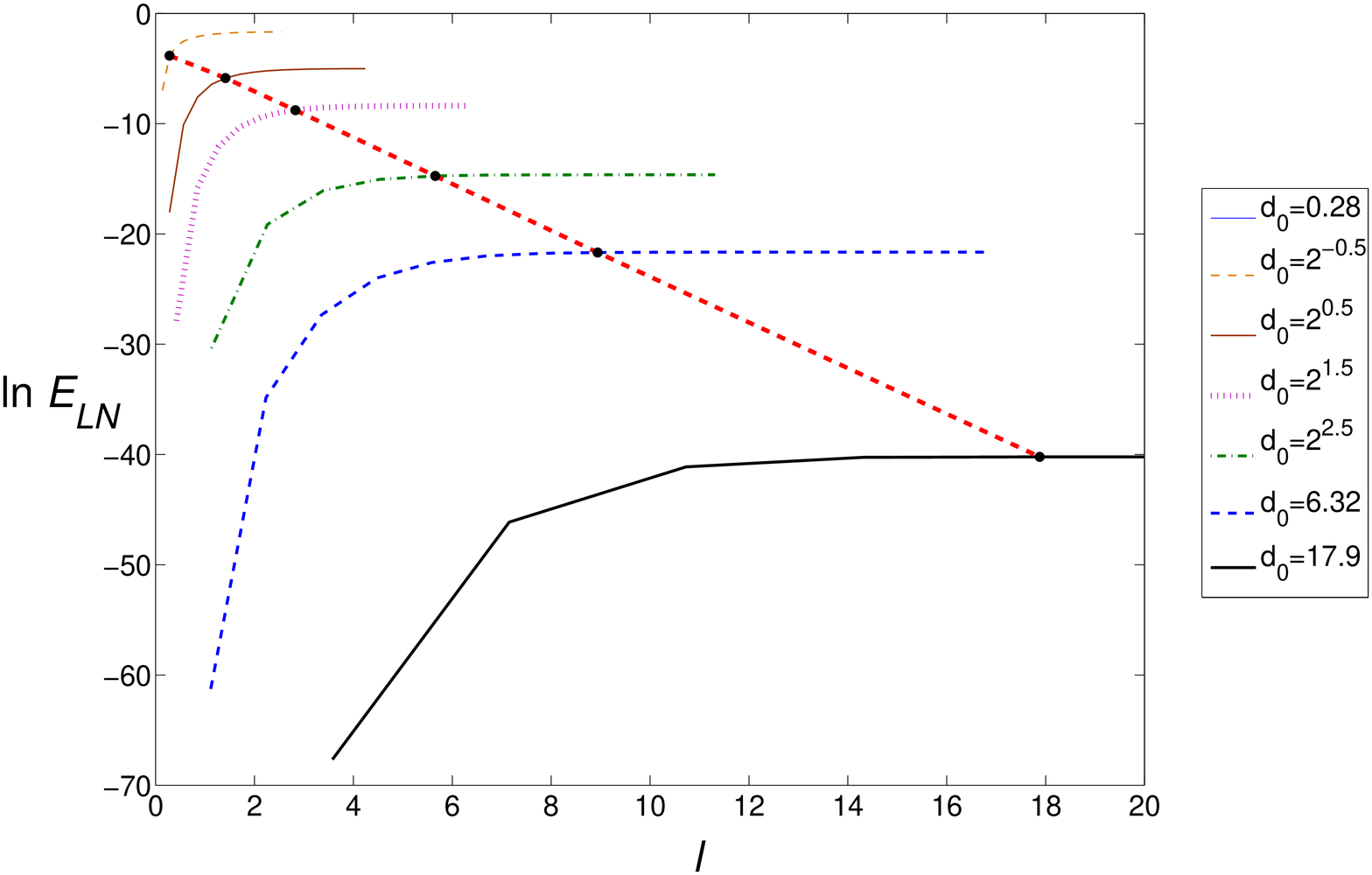}
\caption{(Color online).
Noncritical chain. $\ln E_{LN}$ as a function of $l$ for various values of $d_0$.
The points $E_{LN}(l=d_0)$ fit a linear curve (broken red) with slope $\sim -2.1$.
}
\label{non1}
}
\end{figure}

We now turn to investigate long-range entanglement in the
noncritical field, characterized by a length scale $m^{-1}$,
where in units of the particles' spacing in the harmonic chain,
the length scale is $\xi$, defined in Eq. (\ref{Nt}). Due to
the existence of a length scale, entanglement has to be characterized
by two dimensionless parameters $d$, $l$ given in units of
$m^{-1}$: $d=D/\xi$ and $l=L/\xi$. Numerically we have confirmed
that $$E_{LN}(\xi,D,L)=E_{LN}(\xi \, x,D/x,L/x)$$
in the continuum limit, {\it i.e.} when we observe no difference in
$E_{LN}$ as we simultaneously increase $N$ and $\alpha$ such
that $N / \xi$ remains constant. The noncritical regime reduces
to the critical one if we take $d\to0$ and $l\to 0$.

The noncritical regime is characterized by several limits
depending on the size of $l$ and $d$ with respect to $1$ and
with respect to each other. We expect that in correspondence
with the von Neumann entropy, in the limit $l>>1$ the scaling
becomes independent of $l$. Interestingly, we observe that
this is indeed true only if also 
$l\gtrsim d$. 
In figure
\ref{non1} we plot $\ln E_{LN}(l)$ for several values of a
constant separation $d_0$. We observe that the entanglement
reaches a constant value $E_{LN}^{\rm{sat}}(d_0)$, and thus
distinguishes noncritical systems from critical ones. The
points $E_{LN}(l=d_0)$, which are added for reference, fit
a linear curve (broken red line). Saturation occurs at $l>l_s$,
where $l_s$ is a linear function of $d_0$: $l_s\sim0.75\, d_0+1$.
For small values of $d_0$ saturation is obtained for $l_s\sim 1$
($m=1$ in our notation). Note that each of the curves starts
linearly, showing that the entanglement increases exponentially
with $l$ for small values of $l$.

\begin{figure}[ht]
\center{
\includegraphics[width=3.4in]{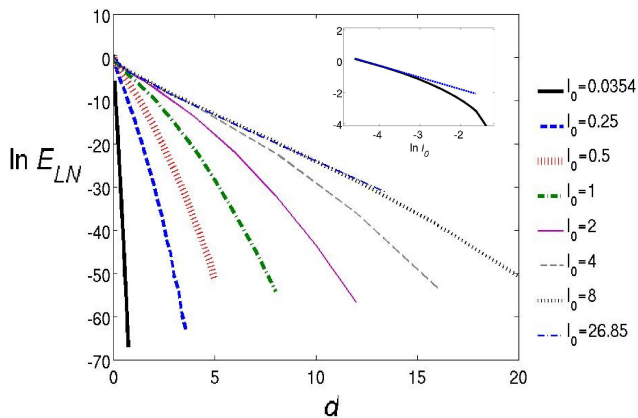}
\caption{(Color online).
Noncritical chain. $\ln E_{LN}$ as a function of $d$ for various values of $l_0$.
We observe exponential decay with a quadratic function of $d$ (which also depends on $l_0$).
In the intermediate saturation regime $l>d>1$ the decay is exponentially linear with $d$.
Inset: power law diverge in the $d\to0$ limit (shown in log-log scale). The power is different
than the $1/3$ in the critical limit and in general depends on $l_0$.}
\label{non_d222}
}
\end{figure}

\begin{figure}[ht]
\center{
\includegraphics[width=3.4in]{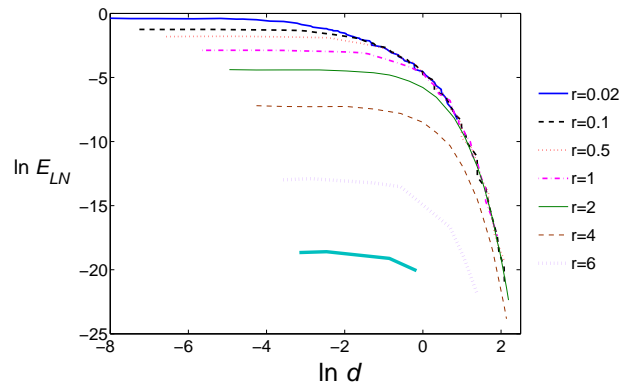}
\caption{(Color online).
Transition from critical to noncritical chain. $\ln E_{LN}$ as a function of $d$ for constant
values of $r=d/l$ ($l$ increases with $d$). Note that all curves
begin with critical behavior as the entanglement is approximately
constant. As $l$ becomes close to $m^{-1}=1$ the noncritical behaviour
emerges. All curves with $r<1$ ($d<l$) coincide
in correspondence with the saturation regime, which is independent
of $l$. The curves with $d>l$ do not coincide and characterize the
regime where $E_{LN}\sim \exp\left(-\beta_{nc}(l)\,d^2\right)$.}
\label{noncr1}
}
\end{figure}

In the opposite limit $d>l$ we observe exponential decay as in
the critical regime, but now with different exponent,
$E_{LN}\sim \exp\left(-\beta_{nc}(l) d^2\right)$, as can be seen in fig. \ref{non_d222},
where $E_{LN}(d)$ is shown for several values of the constant blocks size
$l_0$. Note that in the $d\to0$ limit the entanglement
diverges as a power law, as can be seen in the inset in a log-log
scale. We obtain that the power is different from the $\alpha=\frac{1}{3}$
in the critical limit and in general depends on $l_0$. In addition,
we observe that for $l_0\gg d>1$ the exponential decay becomes linear
with $d$ again, corresponding to the intermediate saturation regime.

In order to observe the transition from critical to noncritical
behavior we plot in figure \ref{noncr1} $\ln{E_{LN}}$ as a function
of $d$ for constant ratios $r=d/l$. Note that all curves begin with
a critical behavior as the entanglement is approximately constant.
As $l$ approaches $m^{-1}=1$ the noncritical behavior emerges
and entanglement starts to decay. It can be seen that all curves
with $r<1$ ($d<l$) coincide at a certain point. This corresponds to
the saturation regime, in which the entanglement decays as
$E_{LN}\sim \exp(-2.25 d)$, independent of $l$. The curves with $d>l$
do not coincide and characterize the regime where $E_{LN}\sim \exp\left(-\beta_{nc}(l)\, d^2\right)$.


We would like to conclude with our main results. The logarithmic
negativity, which is a genuine measure of entanglement, is finite
in the continuum limit. It is distinguished from the classical
correlations especially in the critical limit, where it decays
exponentially with the separation, while classical correlations
decay as a power law. As the blocks approach each other, the
entanglement diverges as a power law, where the power seems to
be equal to the universal prefactor of the logarithmic scaling
of the von Neumann entropy of a large block. It would be interesting
to determine analytically, whether methods from conformal field
theory may be applied to the negativity and obtain $\alpha=\bar{c}/3$.
We note that much like the entropy of entanglement of a single block, the
scaling of long-range entanglement allows us to discriminate
critical from noncritical behaviour. Finally, we point out that
for the critical field both logarithmic negativity and mutual
information are scale invariant and depend only on the ratio
between the distance and length of the blocks.

{\em Note added.} --- When finalizing this paper we became aware
of the independent work on long-range entanglement in critical
spin-chains drawing similar conclusions \cite{sugato}.

{\em Acknowledgements --}
This work was supported by the EU Integrated project (QAP) and the
EPSRC QIP-IRC. B.R. would like to acknowledge the Israel science
foundation grant no. 784/06 and German-Israeli foundation Grant
no. I-857. A.R. acknowledges the support of  EPSRC project number
EP/E045049/1. M.B.P. acknowledges support by the Royal Society.

\end{document}